**Begging with a Purpose? Testing Behavioural Hallmarks of First-Order Intentionality in Free-ranging Hanuman Langurs**

Dishari Dasgupta[1], Shriparna Chattopadhyay[1], Sruti Banerjee[1], Pratyusha Adhikary[1], Akash Dutta[2], Manabi Paul[2*] and Anindita Bhadra[1*]

**Author Affiliations**

[1] Dog Lab, Department of Biological Science, Indian Institute of Science Education and Research (IISER) Kolkata, India

[2] Department of Environmental Science, University of Calcutta

**Orcid ID*:***

Dishari Dasgupta: 0000-0001-7163-7672

Shriparna Chattopadhyay: 0009-0005-0264-690X

Manabi Paul: 0000-0002-8183-8299

Anindita Bhadra: 0000-0002-3717-9732

**Acknowledgements**

DD acknowledges financial support from The International Primatological Society (IPS) for their grant support and the Prime Minister's Research Fellowship (PMRF) for sustaining her during the work tenure. The authors are grateful to Mr Tuhin Subhra Pal for preparing the field site map.

**Funding**

This work was funded by the small research grant award received from The International Primatological Society (IPS). This work was also supported by the Department of Biological Science, Indian Institute of Science Education and Research (IISER), Kolkata, India and Prime Ministers Research fellowship, Ministry of Education, Government of India.

**Declarations:**

**Ethical Approval**

This study was based on non-invasive behavioural observations of free-ranging Hanuman langurs and did not involve any handling or physical interference with the animals. All procedures complied with the ethical guidelines prescribed by the Indian Institute of Science Education and Research (IISER) Kolkata and were approved by the relevant institutional and local ethics committees (IISER/IEC/2024/01).

**Competing interests**

We have no competing financial interests.

**Authors' contributions**

DD, AB and MP conceptualized the study and designed the fieldwork. DD, SC, SB, PA and AD carried out the fieldwork. DD and SC coded the entire data. DD carried out all the statistical analyses. AB and MP supervised the work.  DD, AB and MP drafted the manuscript.

**Abstract**

Intentional communication has been studied extensively in primates, yet evidence from free-ranging non-ape species remains limited. Human-directed food-solicitation gestures in Hanuman langurs (*Semnopithecus entellus*) have recently been described, but whether these behaviours exhibit behavioural hallmarks associated with first-order intentionality remains unknown. Here, we experimentally investigated the presence of these hallmarks in free-ranging Hanuman langurs across six anthropogenic sites in southern West Bengal, India. We conducted 360 experimental and control trials and quantified behavioural markers commonly used to operationalize intentional communication. Experimental trials elicited audience checking, recipient-directed orientation, rapid approach responses, food-solicitation gestures and gestural flexibility, whereas these behaviours were rare or absent in control trials. Differences between experimental and control conditions were significant across all six study sites. Signalling also ceased following food acquisition, consistent with the stopping rule associated with an Apparently Satisfactory Outcome. Our findings demonstrate the presence of multiple behavioural hallmarks linked to first-order intentionality in the human-directed gestural communication of free-ranging Hanuman langurs. These results extend the study of intentionality beyond apes and provide new insights into the evolutionary distribution of intentionality-related traits across primates.

**Introduction**

The term 'intentionality' derives from the Latin in-tendere, meaning 'to aim at' or 'to direct toward', and was developed in medieval scholastic philosophy before being re-introduced in the 19th century by Franz Brentano (Rodrigues & Fröhlich, 2024; Haugh & Jaszczolt, 2012). Brentano argued that mental phenomena was defined by intentionality where beliefs, desires, and perceptions was always about something (Brentano, 1874; Graham et al.,2019). This distinguishes goal-directed behaviour from automatic responses to stimuli (Townsen et al.,2017). Therefore, a signal is intentional if produced to influence others; if it's a fixed or reflexive output, it is not. Paul Grice later built on this, framing communication as a process of recognizing speaker intentions (Grice, 1957).

According to Grice, a communicative act requires that a signaller, who intends to influence a recipient's behaviour, intends the recipient to recognise that intention and generate the desired response based on recognition (Scott-Phillips,2015). This approach treats communication as requiring the recognition of each other's intentions, often described as "mind-reading." While important in human language research, the Gricean framework is difficult to apply to other species, as the underlying intentions it assumes cannot be directly tested in non-human animals. This creates a methodological limitation in assessing intentionality empirically in non-human animals (Ben Mocha & Burkart,2021; Rodrigues & Fröhlich, 2024). To address this, Dennet developed a new approach which shifted focus from mentalistic definitions and placed it on observable criteria instead. In this approach, a behaviour was considered to be intentional when it was goal-directed, flexible and sensitive to context, without requiring direct evidence of mental states (Dennett, 1983). These hallmarks of intentional communication were adapted from studies of pre-linguistic human infants, where intentionality (illocutionary acts) emerges before coherent speech (Bates et al.,1975; Schel et al.,2022; Rodrigues & Fröhlich, 2024). These criteria allow researchers to evaluate whether communicative signals are stereotypical responses or are produced strategically, directed toward achieving an outcome or a specific goal (Ben Mocha & Burkart,2021; Schell et al.,2022).

Empirical application of these hallmarks has been concentrated largely on great apes, resulting in a literature that is both extensive and taxonomically biased. Studies on chimpanzees (*Pan troglodytes*; Leavens et al., 2005), bonobos (*Pan paniscus*; Hobaiter & Byrne, 2011), gorillas (*Gorilla gorilla*; Genty et al., 2009), and orangutans (*Pongo* spp.; Cartmill & Byrne, 2007) have demonstrated that these species use gestures flexibly, direct them toward specific recipients, adjust them based on attentional context and persist when initial attempts fail (Byrne et al.,2017; Rodrigues et al., 2021). Such findings have led to the view that gestural communication in apes is intentional and under voluntary control and may represent an evolutionary precursor to human language (Arbib et al.,2008; Tomasello,2008). However, the dominance of ape-based studies limits broader inferences about the evolutionary distribution of intentional communication. Recent work has begun to extend this framework to non-ape primates, particularly monkeys. Evidence from species such as rhesus macaques (*Macaca mulatta*; Canteloup et al., 2015), olive baboons (*Papio anubis*; Bourjade et al., 2014; Molesti et al.,2020), red-capped mangabeys (*Cercocebus torquatus*; Meunier et al., 2012; Shell et al.,2022), squirrel monkeys (*Saimiri sciureus*; Anderson et al., 2007), and tufted capuchins (*Cebus apella*; Leavens et al., 2010) suggests that hallmarks of intentional communication, including audience sensitivity, persistence and flexible signal use, may be more widespread across the primate lineage than previously assumed.

Despite growing interest in intentional communication, studies on non-ape primates remain relatively limited compared to the extensive work on apes. Moreover, there is considerable variation in how behavioural hallmarks are defined and applied, making cross-species comparisons difficult (Rodrigues & Fröhlich, 2024). In addition, much of the existing research has been conducted in captive or semi-controlled settings, limiting our understanding of how intentional communication operates in free-ranging conditions. These limitations are particularly evident in the Indian context, where studies on intentional communication are scarce. Among the non-ape monkeys found in India, bonnet macaques (*Macaca radiata*) have received some attention, with studies demonstrating goal-directed communicative behaviours in both social and human-interactive contexts (Deshpande et al.,2018; Gupta & Sinha, 2018). However, systematic investigations of intentional communication remain sparse. This gap is especially evident in the case of Hanuman langurs (*Semnopithecus entellus*), a widely distributed Old World monkey frequently found in human-modified habitats. Despite their ecological flexibility and close association with humans, langurs have been largely absent from the intentionality literature.

Recent work by Dasgupta et al. (2025) has documented food-solicitation gestures (here onwards these gestures will be referred as 'begging') in free-ranging Hanuman langurs ranging in and around the Dakshineswar temple complex, West Bengal, India. This study shows that langurs use a repertoire of human-directed gestures, including tactile and postural signals, initiated to obtain food from humans. Such behaviours are context-dependent, socially directed and often successful, suggesting that they may function as structured communicative acts rather than simple conditioned responses. However, whether these begging gestures meet the established behavioural criteria for intentional communication remains unclear. Furthermore, existing observations are largely restricted to a single site, raising the possibility that these behaviours may reflect localised adaptations rather than general features of langur communication.

The present study builds on this work by systematically examining the presence of intentional hallmarks in the human-directed gestural communication of free-ranging Hanuman langurs across multiple sites in southern West Bengal by applying the operational framework of intentionality (Rodrigues & Fröhlich, 2024). Here we ask whether begging gestures exhibit behavioural markers consistent with intentional communication and whether such patterns are shared across troops located in other sites.

## Methods

Study Sites and Subjects

The study was conducted on six troops of free-ranging Hanuman langurs across southern West Bengal, India, located in Sarenga (SA, 22.53 N & 88.19 E), Nangi(NG , 22.50 N  & 88.21 E), Dakshineswar (DG, 22.66 N & 88.36 E), Chandannagore (CD, 22.87 N & 88.38 E), Kalyani (KA, 22,98 N & 88.43 E), and Habibpur (HA, 23.18 N & 88.52 E). These sites represent human-modified, anthropogenic landscapes where langurs frequently interact with people and partially rely on human food provisioning (Fig. 1). Observations were carried out on adult and subadult individuals across all six troops. Focal langurs were selected opportunistically based on proximity (approx. 3–5 m from the observer) and visibility. To minimize pseudo-replication, each individual was tested only once per day, and subjects were never used in both experimental and control conditions within the same 24-hour period.

Commented [AB1]: Here, give the abbreviations within brackets for each site

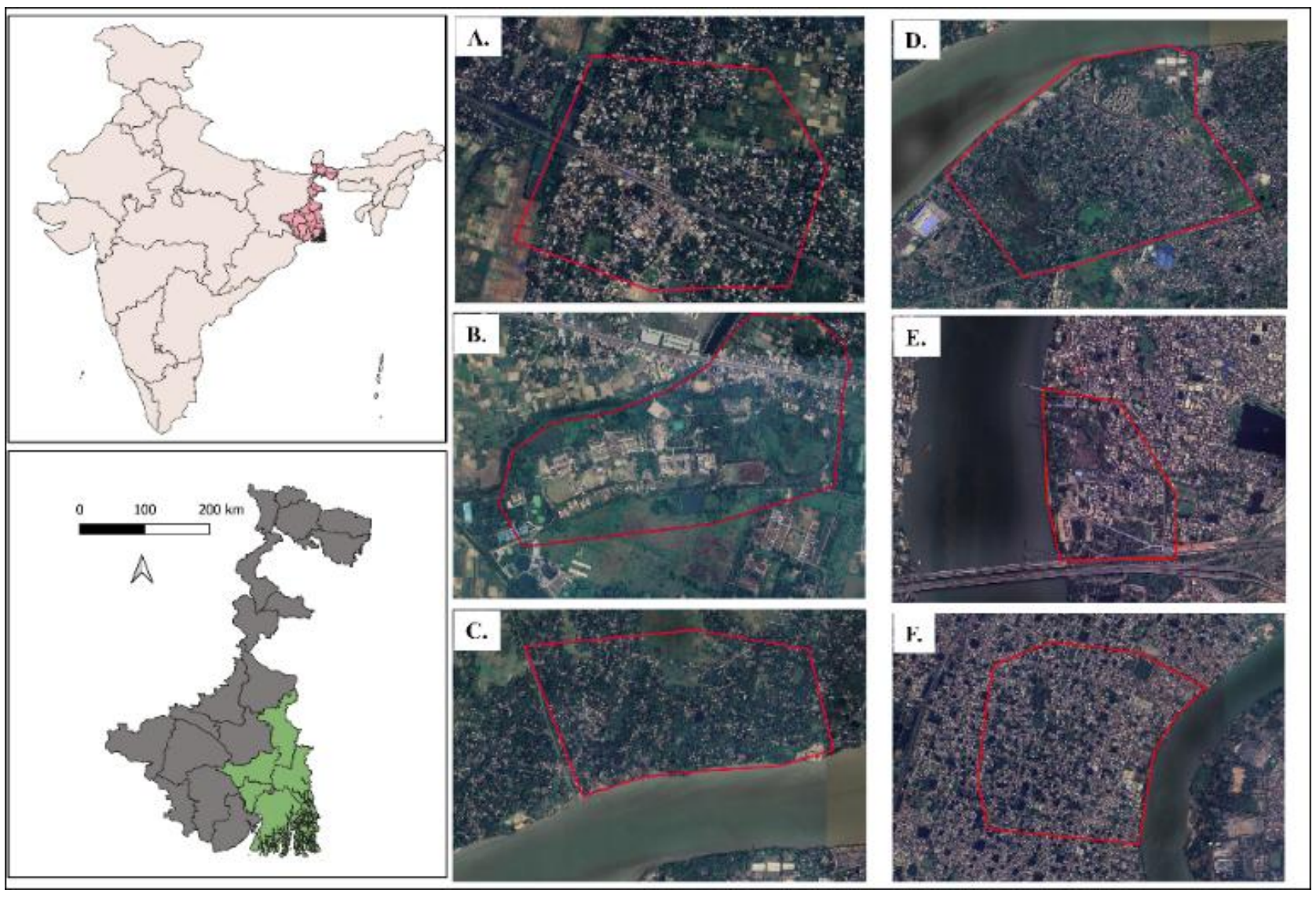

Figure 1: Field site map of the six study sites located in the southern part of the state of West Bengal, India. The study sites are as follows: A. Habibpur,  B. Kalyani,  C. Sarenga D. Nungi E.Dakshineswar  F. Chandannagar

A total of 360 trials were conducted, consisting of 180 trials separately for experimental and control conditions (30 trials per condition, per site). Following established protocols for testing the presence of hallmarks for first-order intentionality, experimental and control trials were conducted on separate days with a minimum interval of five days between consecutive field visits to the same site and between conditions (experiment/control) to reduce the likelihood of short-term associative learning (Leavens et al., 2005; Anderson et al., 2010) (ESM 1: Video of the Experiment and Control).

Experimental Trials

The experimenter visibly displayed a piece of bread while raising one arm for 10 seconds (Cue 1), followed by a 10-second concealment period. If the focal langur produced a begging gesture, the food was delivered (Fig 2).

Since the focal subjects were free-ranging langurs living in close proximity to humans and largely thrive on human-offered food items (Dasgupta et al., 2021) the food reward was delivered immediately upon gesture production. This protocol was strictly maintained for safety reasons. Any deliberate pause or delay in delivering the reward (standard 'response-waiting' and 'persistence' paradigm) could have elicited frustrated or aggressive behaviours from the langurs, posing a significant physical risk to the experimenter. Immediate delivery served as the Apparently Satisfactory Outcome (ASO), confirming the attainment of the signaller's goal (Hobaiter and Byrne, 2014; Townsend et al., 2017). If no response was observed after cue 1, the stimulus (raised arm + holding a bread) was repeated for an additional 10 seconds (Cue 2) to offer the focal langur another chance to respond.

Control Trials

Control trials followed an identical sequence and timing with the experimental trials, except that the experimenter displayed an empty hand and no food was delivered if the focal langur approached the experimenter. This condition was used to assess selective production by determining whether gestures were voluntary social acts produced only when a goal was attainable, or merely reflexive responses to human movement or presence (Leavens et al., 2005; Townsend et al., 2017).

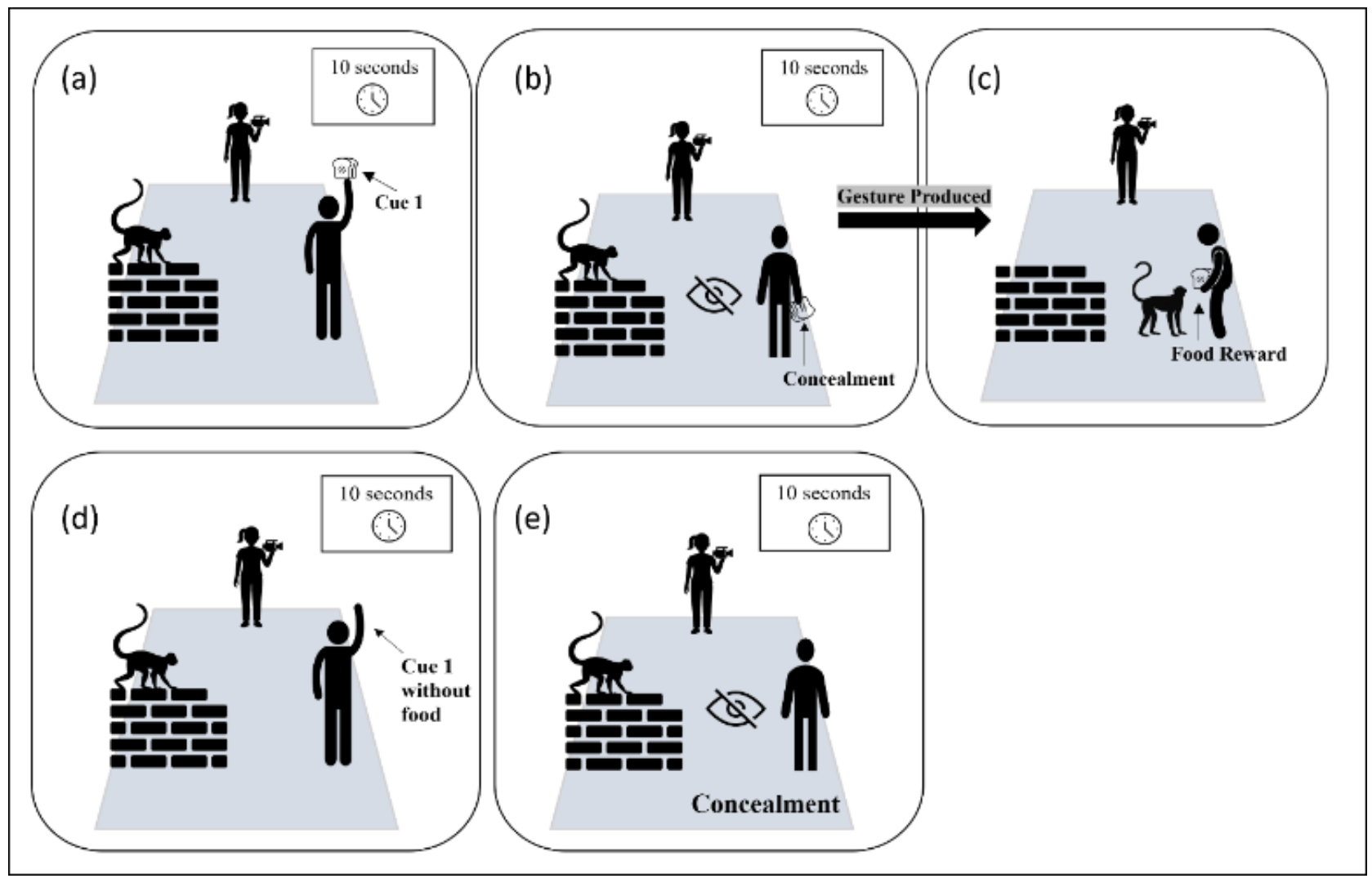

Figure 2. Experimental and control trial protocol for assessing intentional gestural communication in free-ranging Hanuman langurs. (a) Experimental Cue 1: the experimenter raised one arm while visibly displaying bread for 10 s. (b) Concealment phase: the bread was hidden from the langur for 10 s. (c) If the focal langur produced a begging gesture, the food reward was delivered immediately. If no response occurred, the cue sequence was repeated once (Cue 2 , not shown). (d–e) Control trials followed the same sequence and timing, except that the experimenter displayed an empty hand and no food reward was provided.

Behavioural Definitions and Coding

All trials were video-recorded and subsequently analysed. All gestures exhibited by the focal langurs were defined as discrete, mechanically ineffective physical movements directed towards a specific recipient to elicit a voluntary response (Hobaiter and Byrne, 2011; Gupta and Sinha, 2019; Schel et al., 2022). To evaluate the presence of first-order intentionality, the following behavioural hallmarks were coded (Table 1):

| | **Hallmark** | **Definition** | **Operationalization** | **References** |
|---|---|---|---|---|
| 1 | Audience checking | Signaller monitors the recipient to ensure they are a suitable communicative partner before initiating an interaction. | Signaller looks at the recipient (LO) immediately before approaching the recipient. Coded as yes/no. | Bates et al., 1975; Rodrigues & Fröhlich, 2024 |
| 2 | Goal-directedness (Desire Criterion) | Signaller's investment of time or effort reflects a desire for a specific outcome. | Four measures were used. | Ben Mocha & Burkart, 2021; Dasgupta et al.,2025 |
| 2a | Look duration | Duration of visual attention directed at the experimenter. | Measured in milliseconds (ms). Time duration between LO start time and end time was measured. | |
| 2b | Approach latency | Time between the experimenter's cue and the onset of the langur's approach. | Measured as time difference between langur's approach time and trial start time. Time measured in ms. | |
| 2c | Begging gestures (BG) | Presence of recipient-directed gestural displays toward the experimenter. | Begging gestures were coded as present/absent during the trial. | |
| 2d | Apparently satisfactory outcome (ASO) | Signalling terminates (BG Stops) immediately after the goal (food acquisition) is achieved. | Cessation of begging gestures after receiving the food. Coded as yes/no. | |
| 3 | Social use (targeted signalling) | Signal is a socially directed act, not a general behavioural display. | Signaller aligns nose-body (NB) orientation toward the recipient during signal production. Coded as yes/no. | Bates et al., 1975; Molesti et al., 2020; Rodrigues & Fröhlich, 2024 |

| 4 | Means-ends dissociation (gestural flexibility) | Use of multiple distinct gesture types interchangeably for a single goal. | Number of different begging gestures used by the focal langur to obtain bread (gestural repertoire). | Call & Tomasello, 2007; Rodrigues & Fröhlich, 2024 |
|---|---|---|---|---|
| 5 | Volitional control (selective production) | The ability of the signaller to selectively produce, withhold, or terminate communicative signals in pursuit of a cognitively represented goal | | Schel et al.,2022; Ben Mocha & Burkart, 2021 |
| 5a | Behaviour change (BC) | Alteration of ongoing behaviour following cue presentation, relative to pre-cue state. | Behaviour noted for both states ( pre-cue and post-cue) Later coded as yes/no for change in behaviour. | |
| 5b | Approach (AP) | Movement toward the experimenter after cue presentation. | Coded as yes/no; indicates active engagement and initiation of interaction. | |

Table 1 : Behavioural Hallmarks of Intentional Gestural Communication.

(1) Audience Checking: We examined whether the focal langur monitored the recipient to ensure that they were a suitable communicative partner before initiating an interaction (Bates et al., 1975). This was operationalized as the signaller looked at the recipient (LO) just before approaching the recipient (Rodrigues and Fröhlich, 2024).

(2) Goal-Directedness (Desire Criterion): Following the framework proposed by Ben Mocha and Burkart (2021), we used three measures to infer goal-directedness:

(i)Look Duration (Signaller Investment): We measured the duration (in milliseconds) for which the focal langur looked at the experimenter. Within the "Desire Criterion," the investment of time for which the signaller looked at the recipient serves as evidence that the specific outcome is desired by the signaller.

(ii)Approach Latency (Signaller Investment): This was calculated as the time difference between the start of the experimenter's cue and the onset of the langur's approach. Here the time investment in terms of speed reflects the "eagerness" or motivation of the focal langur to obtain its goal.

(iii)Begging Gestures (BG): We recorded the presence of recipient-directed gestural displays towards the experimenter, reflecting goal-directed communicative intent (Leavens et al., 2005; Schel et al., 2013).

(iv)Apparently Satisfactory Outcome (Stopping Rule): We examined whether the begging gesture (BG) ceased immediately after food acquisition. A cardinal marker of intentionality is the Stopping Rule, where signalling terminates once an Apparently Satisfactory Outcome (ASO), a reaction that satisfies the signaller's goal is achieved.

(3) Social Use (Targeted Signalling): To determine if the gesture was a socially directed act, we coded the nose-body (NB) orientation of the focal langur (Bates et al., 1975). A signal is classified under Social Use if the signaller aligns their body and nose toward the recipient while communicating the signal (Molesti et al., 2020).

(4) Means-Ends Dissociation (Gestural Flexibility): Means-ends dissociation is satisfied when a signaller uses a variety of different gesture types interchangeably for a single end goal (Bruner, 1981; Call and Tomasello, 2007). We evaluated whether langurs displayed multiple distinct begging gestures (the gestural repertoire: Dasgupta et al,2025; ESM 2 : Gesture Table) to procure the bread.

(5) Volitional Control (Selective Production):

To distinguish intentional signalling from mechanistic responses, we assessed selective production across experimental and control conditions (Townsend et al., 2017; Ben Mocha & Burkart, 2021).

(i) Behaviour Change (BC): We measured whether the focal langur altered its ongoing behaviour following cue presentation relative to its pre-cue state, indicating attentional reorientation.

(ii) Approach (AP): We quantified whether the focal individual moved towards the experimenter after cue presentation, indicating active engagement and initiation of interaction.

Inter-observer Reliability

To assess coding reliability, 20% of the dataset was independently coded by a second observer (Cohen,1960). Inter-observer agreement was high across all variables, as assessed using Cohen's kappa: approach ($\kappa = 0.84$), nose orientation ($\kappa = 0.84$), body orientation ($\kappa = 0.84$), begging presence ($\kappa = 1.00$), food receipt ($\kappa = 1.00$), and gesture type ($\kappa = 0.82$). For the variable 'look', agreement was 100%, resulting in an undefined κ value due to a complete lack of variance in the reliability sample.

Statistics

We used R version 4.4.1 (R Core Team 2024) to conduct all statistical tests. Because individual langurs could not be uniquely identified across trials, we performed analyses at the trial and site level rather than at the individual level, with a sample of 30 experimental trials and 30 control trials per site (N = 360 trials total). All tests were conducted with the alpha level set at 0.05 and Bonferroni adjustments applied for multiple comparisons where noted (adjusted $\alpha = 0.0083$ for six sites). For Audience Checking, we compared the proportion of trials with looking behaviour (LO) between experimental and control conditions at each site using Fisher's exact tests; because the experimental condition showed a ceiling effect (100% occurrence), the odds ratio was infinite. Therefore, we calculated risk differences with Newcombe's hybrid-score confidence intervals. For Goal - Directedness, we assessed Look Duration using Wilcoxon rank-sum tests with Bonferroni adjustment (all adjusted $p < 0.001$) and compared Approach

Latency across experimental trials from six sites using a Kruskal -Wallis test (H = 9.90, df = 5, p = 0.078). Begging Gestures (BG) and Apparently Satisfactory Outcome (Stopping Rule) were analysed descriptively alongwith Fisher's exact tests where applicable. For Social Use (Targeted Signalling) and Volitional Control (Selective Production), including Behaviour Change (BC) and Approach (AP), we applied Fisher's exact tests with Bonferroni correction, reporting risk differences as effect sizes. Finally, for Means - Ends Dissociation (Gestural Flexibility), we descriptively counted gesture types per site and calculated the Shannon diversity index to characterise evenness of gesture use.

**Results**

Hallmark 1: Audience Checking (Looking at the Experimenter - LO)

The occurrence of audience checking (LO) differed significantly between conditions across all sites. In the experimental condition, LO was observed in 100% of trials at every site. Control-condition proportions were substantial, ranging between 63% to 73% across six sites (Fig 3). Site-wise Fisher exact tests revealed statistically significant differences at all sites after Bonferroni correction (adjusted α = 0.0083): CD (Risk Difference = 30.0 pp, 95% CI [13.5, 46.5], p = 0.0010), DG (36.7 pp, 95% CI [20.0, 53.0], p < 0.0001), HA (26.7 pp, 95% CI [10.3, 43.1], p = 0.0024), KA (26.7 pp, 95% CI [10.3, 43.1], p = 0.0024), NG (36.7 pp, 95% CI [20.0, 53.0], p < 0.0001), and SA (36.7 pp, 95% CI [20.0, 53.0], p < 0.0001). Because the experimental condition exhibited a ceiling effect (100% LO), the associated odds ratios were infinite. Therefore, we used risk differences with Newcombe's hybrid-score confidence intervals to quantify the absolute increase. These large and consistent risk differences (26.7–36.7 percentage points) demonstrate that the experimental condition was strongly associated with a higher proportion of LO trials across all sites.

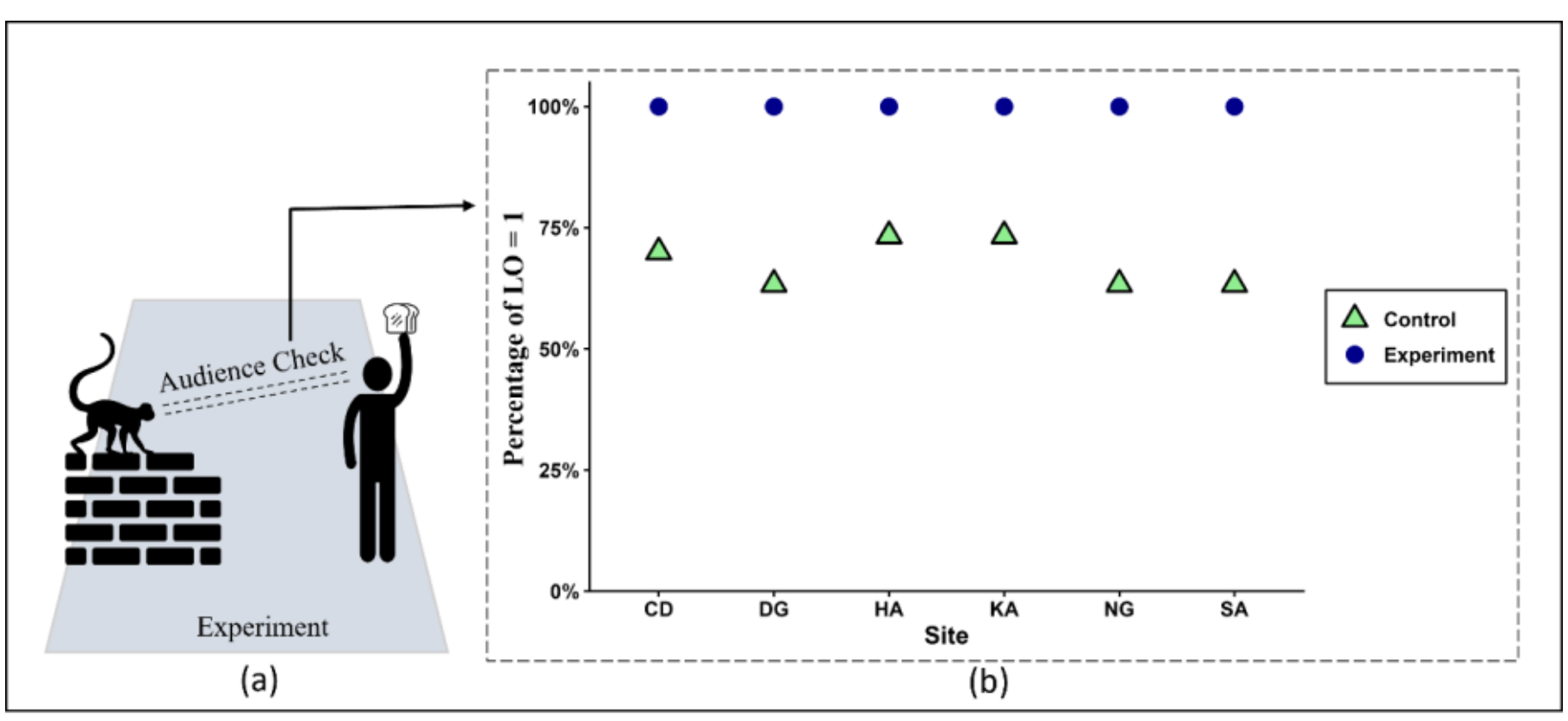

**Figure 3**: Hallmark 1: Audience checking. (a) Operationalization of the hallmark. (b) Percentage of “look” events (LO = 1) exhibited by the focal langur during experiment (dark blue circle) versus control (light green triangle) trials across all six field sites. Site abbreviation mentioned in ‘Methods’ section.

Hallmark 2: Goal-Directedness (Desire Criterion)

(i) Looking Duration (LO Duration)

Looking duration was consistently higher in the experimental condition than in the control across all sites (Wilcoxon rank-sum Test : all Bonferroni-adjusted $p < 0.001$).While control trials were dominated by zero durations, experimental trials showed sustained looking behaviour. Site-wise comparisons were significant (Fig 4i)

(ii) Approach Latency

Approach latency varied across sites within the experimental condition (Fig 4 ii). Median values ranged from 503 ms (NG) to 2913 ms (SAR). Latency distributions were right-skewed across all sites, with most approaches occurring rapidly and occasional delayed responses contributing to higher means. The modal latency values around zero ms indicates immediate or near-immediate responses in a subset of trials. Despite variability in mean latency, median values were broadly comparable across sites, suggesting similar central tendencies across locations. (Kruskal–Wallis test : H = 9.90, df = 5, p = 0.078).

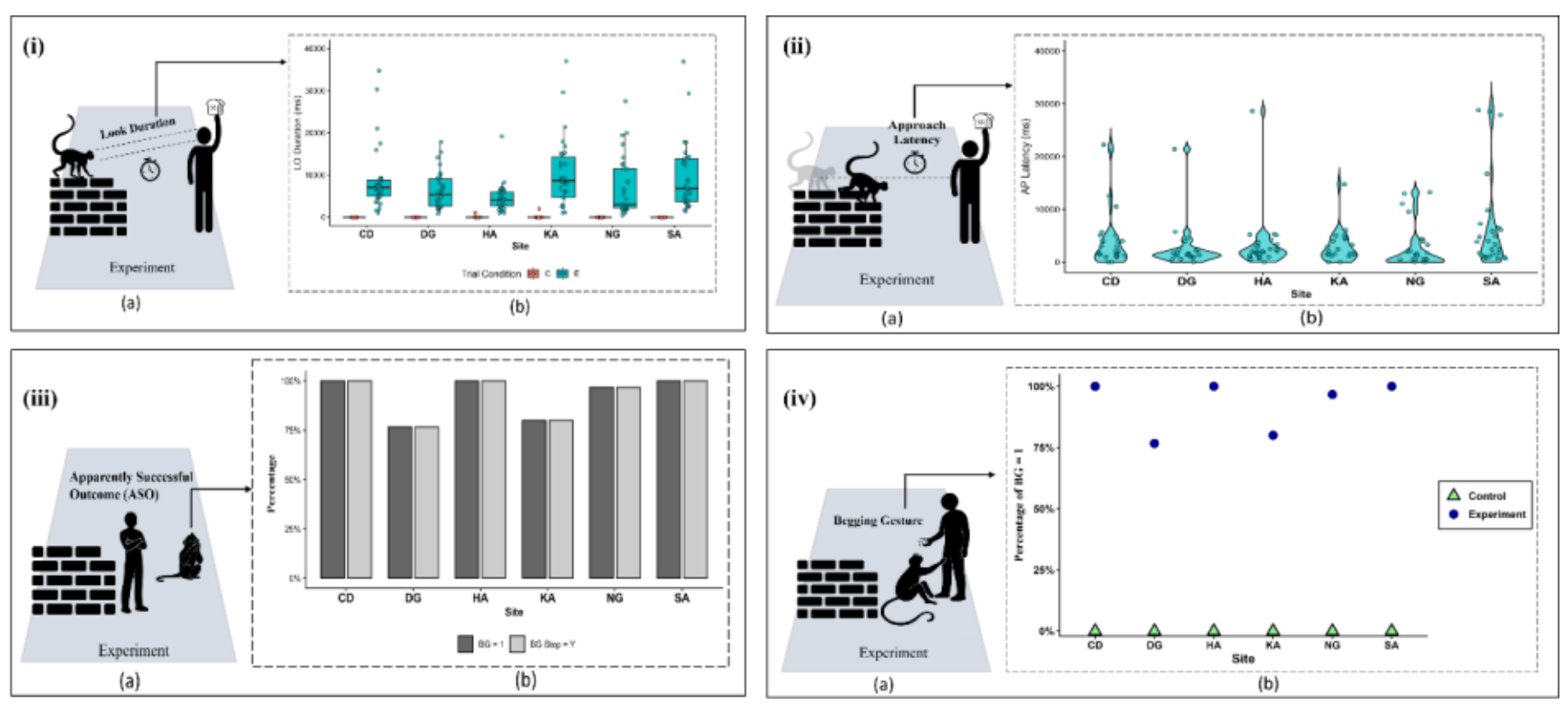

Figure 4. Hallmark 2: Desire Criterion (goal-directedness). (a) Operationalization of the hallmark. (b-i) Box-and-jitter plot of LO (Look) duration for the experimental (E0 and control (C) trials across all six sites. (b-ii) Violin-and-jitter plot of approach (AP) latency during experimental trials across all six sites. (b–iii) Bar plot showing the percentage of ASO (BG Stop) events after the focal langur received bread following a begging gesture (BG = 1) in experimental trials, for each site. (b-iv) Percentage of "begging" events (BG = 1) in experimental (dark blue circles) versus control (light green triangles) trials across all six sites. Site abbreviations are given in the Methods section.

(iii) Apparently Satisfactory Outcome (BG Stop after Food)

Across all sites, individuals consistently ceased begging behaviour following food acquisition. The proportion of cessation was uniform at every site (Site: 6, N: 30, Proportion for each site:1; Fig 4 iii)

(iv) Begging Behaviour (BG)

Begging was absent in control trials across sites but occurred frequently in the experimental condition (76.7–100%; Fig 4 iv). All comparisons were significant (all Bonferroni-adjusted p < 0.001), with large effect sizes (Risk Difference: 0.77–1.00).

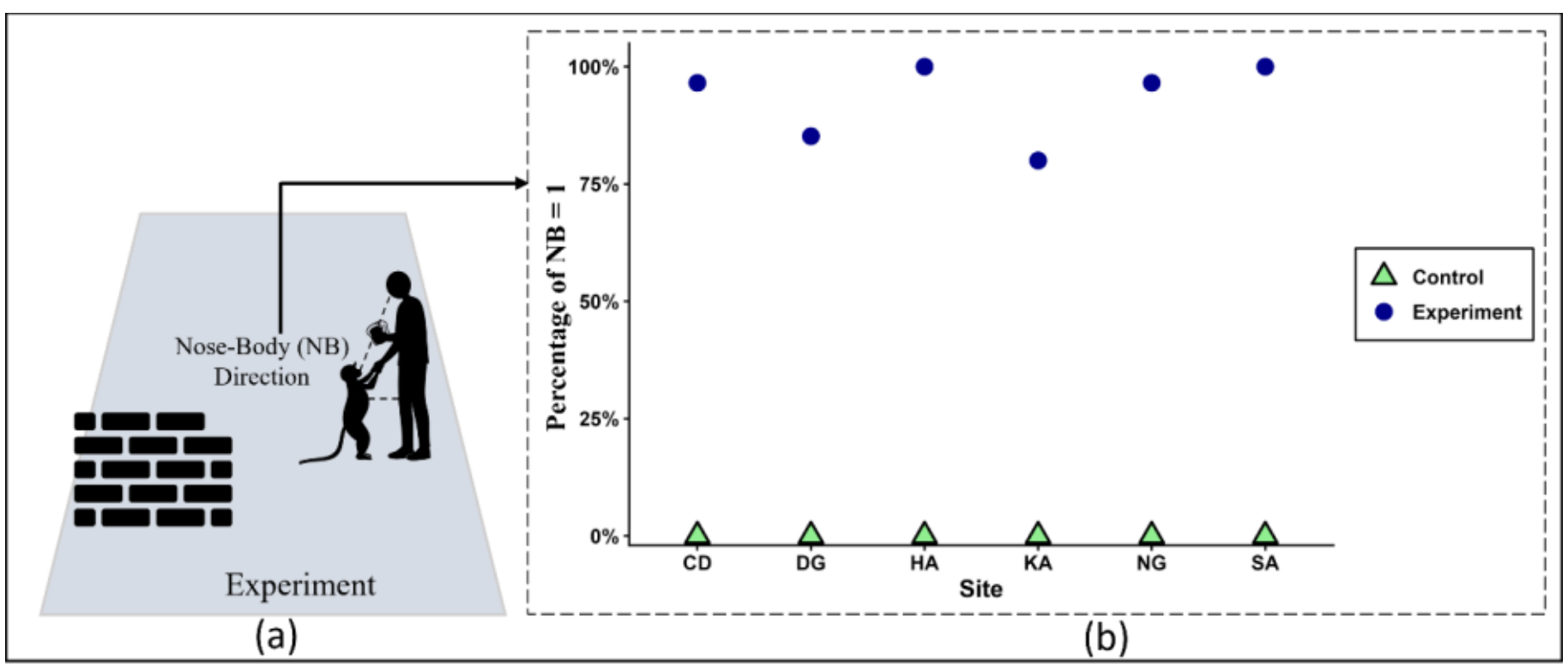

Figure 5. Hallmark 3 : Social Use. (a) Operationalization of the hallmark. (b) Percentage of events during which the focal langur's nose and body were directed towards the experimenter (NB = 1) during experimental (dark blue circle) versus control (light green triangle) trials across all six field sites. Site abbreviation mentioned in Methods section.

Hallmark 3: Social Use (Targeted Signalling)

Nose–Body Direction (NB) responses were absent in control trials across all sites but occurred frequently in the experimental condition (0.80–1.00). Differences between conditions were significant at all sites (Fisher Exact Test: all Bonferroni-adjusted p < 0.001), with large effect sizes (Risk Difference: 0.85 to 1) (Fig 5).

**Hallmark 4: Volitional Control**

(a) Approach Behaviour

Approach responses were hardly present (0 to 3.33%) in control trials but occurred frequently in the experimental condition (76.7–100%). Differences between conditions were significant at all sites (all Bonferroni-adjusted p < 0.001), with large effect sizes (risk differences: 0.77–1.00) (Fig 6).

(v) Behaviour Change (BC)

Behaviour change was rare or absent in control trials (0–3.33%) but occurred frequently in the experimental condition (0.73–0.97). Differences between conditions were significant across all sites (all Bonferroni-adjusted p < 0.001), with large effect sizes (Risk Difference: 0.73–0.97). Comparisons across these behaviours largely exhibited complete or near-complete separation, indicating extremely strong condition dependence. Overall, behaviours reflecting volitional control were consistently and strongly associated with the experimental condition across all sites (Fig 6).

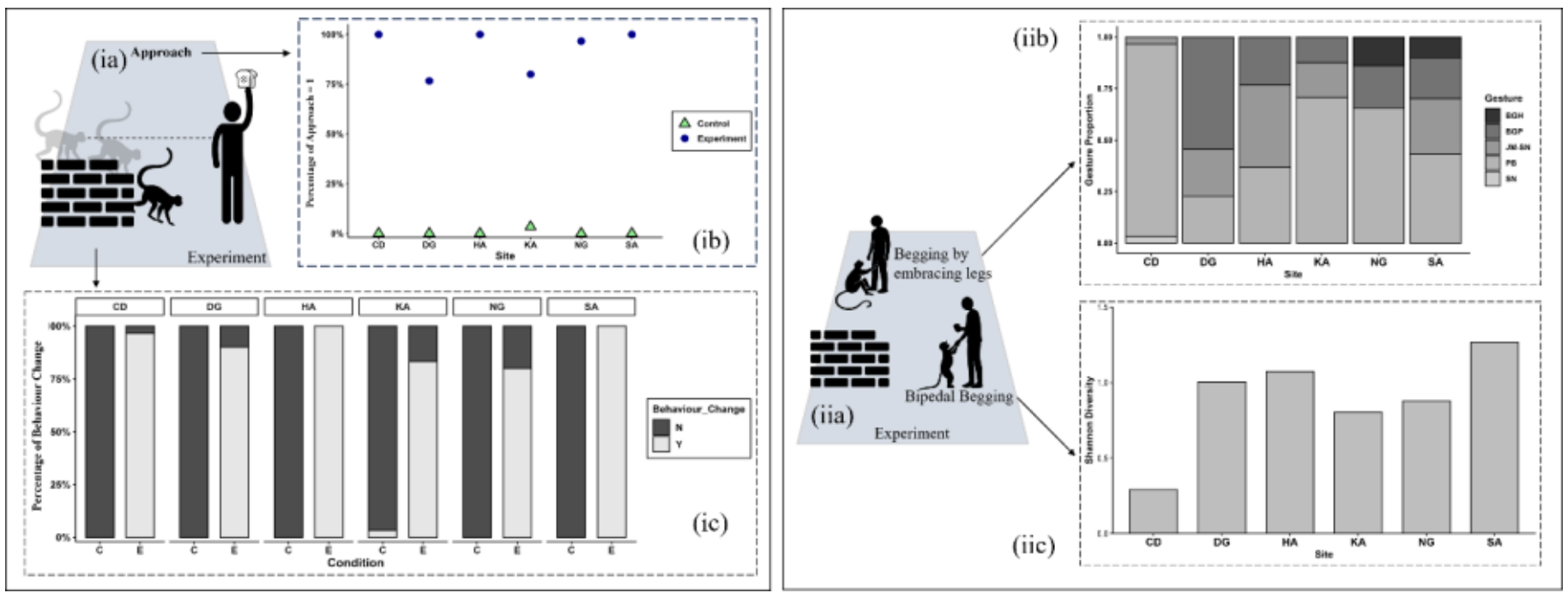

Figure 6: Hallmark – Volitional control (behaviour change after cue).(ia & iia) Operationalization of hallmark. (ib) Percentage of Approach events (AP = 1) by focal langurs during experimental (dark blue circle) versus control (light green triangle) trials across all six field sites. (ic) Stacked bar plots showing the proportion of behavioural changes exhibited by focal langurs after receiving the cu, for experimental (E) and control (C) trials at each sites. (iib) Stacked bar plot of the proportion of different gesture types produced during experimental

trials across the six sites. (iic) Bar plot of the Shannon diversity index of gestures produced during experimental trials at each site. Site abbreviations are given in Methods section.

Hallmark 5: Means–End Dissociation (Gestural Flexibility)

Focal langurs at all sites exhibited multiple gesture types, with repertoire richness ranging from three to four gestures. A total of five gesture types were recorded (PB, BGP, JM-SN, BGH, SN), with PB being the most frequent overall. Despite similar repertoire richness, gesture distributions differed across sites. CDNR showed strong dominance of a single gesture (PB: 93.3%; Shannon = 0.29), indicating low evenness. In contrast, DG (Shannon = 1.00), HAB (1.07), and SAR (1.27) exhibited more even distributions. KAL (0.80) and NG (0.88) showed intermediate patterns. Overall, while gesture repertoires were broadly similar across sites, the distribution of gesture use varied, suggesting flexible use of these gestures(Fig 6).

**Discussion**

This study investigated whether the established behavioural hallmarks of first-order intentionality, defined as voluntary signalling in pursuit of a cognitively represented goal, are present or not in the human-directed gestural communication of free-ranging Hanuman langurs living in anthropogenic landscapes across multiple sites in southern West Bengal (Dennet,1983; Townsend et al.,2017; Ben Mocha & Burkart,2021). Our results provide evidence of presence of intentionality hallmarks among these free-ranging colobine species. In the experimental condition, langurs displayed Audience Checking (Figure 2) by visually monitoring the experimenter in nearly all trials, approached and begged at consistently high rates (Volition Control – Figure 7 & 8) and visually oriented themselves towards the experimenter during gestural communication (Social Use through targeted signalling – Figure 6) (Schell et al.,2022; Gupta & Sinha,2019).They also displayed sustained looking durations and eagerness to approach the experimenter, satisfying the Desire Criterion of Goal-directedness (Figure 3,4,5) (Ben Movcha & Burkart, 2021).

In contrast, the control condition (empty hand) elicited almost none of these classic goal-directed behaviours. If gesturing were merely a learned association between the experimenter's presence and a food reward, we would expect langurs to approach and beg in both conditions, because the experimenter appears in both with their raised hand cue. The near-complete absence of response in control trials demonstrates that the subjects were not simply conditioned to the human experimenter or the hand raising cue (Leavens et al.,2005; Anderson et al.,2010). This stark dissociation offers the first line of evidence against a simple zero-order or associative learning account. Similar patterns have also been found in ape studies that ruled out simple associative learning by demonstrating that signal production was selective and context-dependent (Leavens et al., 2004, 2019).

While the response to the food cue was quite clear, no approach or begging was observed in a minority of experimental trials at several sites, despite the clear visual cue of the food (DG:7/30; NG: 1/30; KAL: 6/30; Figure 7 & 8). This was not a failure of the manipulation but likely reflects a genuine decision on the part of the free-ranging langur. In these highly provisioned anthropogenic sites, some individuals may simply have been satiated at the time of testing. The presence of these 'skip' trials paradoxically strengthens the intentionality interpretation: gesturing was not an automatic reflex triggered by food, but a flexible behaviour that could be withheld when the internal motivational state (satiation) rendered the goal temporarily not worth pursuing (Cartmill & Byrne, 2010). Thus, responding only when food is present, but not always even then, constitute strong behavioural evidence that the response is goal-directed and decision-based, rather than a rigid stimulus-output rule.

Audience checking and the orientation of the signaller toward the recipient, hallmarks that have been validated in non-human primates (Call & Tomasello, 2007 and Leavens et al., 2019), was also found in our experimental trials. Notably these behaviours were virtually absent in control trials. Goal-directed communication was reflected in the rapid approach latencies (median latencies as low as 503 ms, with a modal latency of 0 ms at several sites) and in the sustained looking duration that characterised the experimental condition. These temporal measures indicate that the food was a cognitively represented goal and that the langurs were genuinely invested in obtaining it, a concept consistent with the "desire criterion" (Ben Mocha & Burkart, 2021).

Means-end dissociation, the use of different gestures (the means) for the same goal (food) was clearly evident across all our study sites, with langurs employing up to four distinct gesture types (Bruner,1981; Call and Tomasello,2007). Interestingly, flexibility varied across sites: some troops showed diverse, evenly distributed repertoires (Shannon indices > 1.0), whereas others relied more heavily on a single gesture (Figure 10,11). Regarding the 'stopping rule', for every single trial where a langur had begged and subsequently received food, the begging behaviour ceased immediately (proportion = 1.00 across all sites). This unambiguous termination of signalling once an 'Apparently Satisfactory Outcome' (ASO) has been achieved, distinguishes their communication from being an arousal-driven reflex (Townsen et al.,2017; Hobaiter and Byrne,2014).

Our findings also align with and extend the growing body of literature on gestural communication in non-ape primates, which has begun to demonstrate that intentionality is not restricted to the hominid lineage. Detailed studies on food-requesting in wild bonnet macaques (*Macaca radiata*) have previously demonstrated that Old World monkeys can exhibit hallmarks such as audience checking, persistence, and sensitivity to the recipient's attentional state (Gupta & Sinha, 2018). Similarly, controlled experiments using food-requesting paradigms have shown that olive baboons (*Papio anubis*) can adjust their gestural modality (visual vs. auditory) to match the experimenter's visual attention, a sophisticated form of audience checking (Meunier et al., 2013). Comparable results have also been reported in red-capped mangabeys (*Cercocebus torquatus*), where researchers demonstrated goal persistence and means-end dissociation in conspecific interactions (Canteloup et al., 2020; Schell et al., 2022). Nevertheless, systematic investigations of intentionality in colobine monkeys have remained rare; our study fills this taxonomic gap by extending the intentionality literature beyond cercopithecines (macaques, baboons) to a colobine species.

Moreover, by conducting this study in highly anthropogenic, provisioned environments, we show that such experimental research can be carried out in real-world settings where primates and humans frequently interact. However, it is equally important to acknowledge the intentionality hallmarks we could not test in this study, due to the inherent and often unavoidable constraints of working with free-ranging langurs in an anthropogenic landscape. Four classic hallmarks remain unresolved. First, the Response Waiting hallmark where the signaller pauses after a gesture to await a response, could not be measured (Leavens et al., 2005). In our rapid-testing paradigm, the experimenter provided food almost immediately upon gesturing to ensure safety, as any delay could have elicited aggressive behaviour from the langurs towards the experimenter. Second, persistence across failed attempts, defined as repetition or modification of gestures when the initial signal does not succeed), could not be assessed. Relevant to this is the recent naturalistic study on Hanuman langurs in an urban habitat, which observed that when individuals did not receive a preferred food item, they persistently continued their begging gesture until the desired food was offered, a clear demonstration of persistence (Dasgupta et al., 2025). In our experimental paradigm, deliberately withholding food to induce such persistence could easily escalate the interaction, placing the experimenter at risk from the free-ranging langurs. Third, gaze alternation, the act of rapidly looking between the goal (here, the food) and the experimenter as a marker of triadic communication and joint attention, represents another hallmark that we did not include in our study (Bard,1992; Leavens and Hopkins,1998). Logistically, we could not record this behaviour reliably with a handycam held by the person assisting the experimenter standing at a distance. Finally, elaboration, the addition of novel gestures or a switch to a different signalling modality when the initial gesture fails, could not be safely tested. Resulting elaboration could provoke unpredictable and aggressive behaviour by the langurs which would put the experimenter at risk. Moreover, because langurs are not individually identified, we could not track whether the same individual appeared in multiple trials, precluding analysis of within-subject consistency or learning over time and making it impossible to fully control for pseudo-replication, although the site-level replication and large number of trials mitigate this concern. The continuous presence of human activity, noise, and other disturbances in these environments introduces uncontrolled factors that are absent from captive settings (Liebal et al.,2013). These are not weaknesses of the study per se, but rather necessary trade-offs in moving from controlled laboratory conditions to real-world, free-ranging contexts, and we present them as explicit lacunae to help guide the design of future research.

Despite these lacunae, the present study has demonstrated that several of the established behavioural hallmarks of first-order intentionality are present in the gestural communication of free-ranging Hanuman langurs. The inclusion of a control condition allows us to rule out simple associative learning and reflexive accounts. Our findings suggest that the cognitive capacities underlying intentional communication are more widely distributed across the primate order than previously acknowledged. They further demonstrate that even in ecologically complex, human-modified environments, robust experimental tests of animal cognition are both feasible and informative (Dasgupta et al., 2025; Rodrigues and Fröhlich, 2024). This study demonstrates how Hanuman langurs not only have the ability for intentional gestural communication, but use them in their interactions with humans effectively, highlighting an important adaptation to the Anthropocene in these primates. The hallmarks that could not be tested remain important targets for future research.